\documentstyle[11pt,newpasp,twoside]{article}
\def\edcomment#1{\iffalse\marginpar{\raggedright\sl#1\/}\else\relax\fi}
\reversemarginpar

\begin{document}
\title{The Magellanic Clouds and the Primordial Helium Abundance}
 \author{Manuel Peimbert \& Antonio Peimbert}
\affil{Instituto de Astronom\'{\i}a, Universidad Nacional Aut\'onoma de 
M\'exico. Apdo. Postal 70--264, M\'exico, D.F. 04510}

\begin{abstract}
A new determination of the pregalactic helium abundance based on the Magellanic
Clouds H~II regions is discussed. This determination amounts to $Y_p = 0.2345
\pm 0.0030$ and is compared with those derived from giant extragalactic H~II
regions in systems with extremely low heavy elements content. It is suggested
that the higher primordial value derived by other authors from giant H~II
region complexes could be due to two systematic effects: the presence of
neutral hydrogen inside the helium Str\"omgren sphere and the presence of
temperature variations inside the observed volume.
\end{abstract}

\section{Introduction}
The determination of the pregalactic, or primordial, helium abundance by mass
$Y_p$ is paramount for the study of cosmology, the physics of elementary
particles, and the chemical evolution of galaxies (e. g. Fields \& Olive 1998,
Izotov et al. 1999, Peimbert \& Torres-Peimbert 1999, and references
therein). In this review we briefly discuss the method used to derive $Y_p$ and
its main sources of error as well as a new determination based on observations
of the SMC. This determination is compared with those carried out earlier based
on extremely metal poor extragalactic H~II regions.

The Magellanic Clouds determination of $Y_p$ can have at least four significant
advantages and one disadvantage with respect to those based on distant H~II
region complexes: a) no underlying absorption correction for the helium lines
is needed because the ionizing stars can be excluded from the observing slit,
b) the determination of the helium ionization correction factor can be
estimated by observing different lines of sight of a given H~II region, c) the
accuracy of the determination can be estimated by comparing the results derived
from different points in a given H~II region, d) the electron temperature is
generally smaller than those of metal poorer H~II regions reducing the effect
of collisional excitation from the metastable 2$^3$ S level of He~I, and e) the
disadvantage is that the correction due to the chemical evolution of the SMC is
in general larger than for the other systems.

\section{He$^+$/H$^+$ Determinations}

To derive accurate He$^+$/H$^+$ values we need very accurate $N_e$ (He~II),
$T_e$ (He~II), and $\tau$(3889, He~I) values. Good approximations to determine
He$^+$/H$^+$ for $N_e < 300$ cm$^{-3}$, $13~000$~K~$< T_e < 20~000$~K and
$\tau(3889) = 0.0$, have been presented for the main helium lines by Benjamin,
Skillman and Smits (1999):

\begin{eqnarray}
\frac{N({\rm He}^+)}{N({\rm H}^+)} & = & \frac{I(6678)}{I({\rm H}\beta)}  
2.58 T_4^{0.249 - 2.0 \times 10^{-4} N_e}, \\
& = & \frac{I(4471)}{I({\rm H}\beta)} 2.01 T_4^{0.127 - 4.1 \times 
10^{-4} N_e}, \\
& = & \frac{I(5876)}{I({\rm H}\beta)} 0.735 T_4^{0.230 - 6.3 \times 
10^{-4} N_e}.
\end{eqnarray}

Fortunately for giant extragalacic H~II regions $\tau(3889)$ is very small and
frequently close to zero. $\tau(3889)$ can be estimated together with $N_e$
(He~II) from the 3889/4471 and the 7065/4471 ratios computed by Robbins (1968).

Most authors assume that $T_e$(O~III) = $T_e$(He~II), there are two reasons for
this assumption the He$^+$ and O$^{++}$ emission regions occupy similar volumes
(but not identical ones) and $T_e$(O~III) is easy to measure. Nevertheless the
assumption is correct only for isothermal nebulae, and not for real nebulae if
very accurate abundances are needed. In the presence of temperature variations
along the line of sight the [O~III] lines originate preferentially in the high
temperature zones and the helium and hydrogen lines in the low temperature
zones (e.g. Peimbert 1967, 1995).  It can be shown that the temperature derived
from the ratio of the Balmer continuum to a Balmer emission line, $T_e$(Bac),
is similar to $T_e$(He~II), and that both temperatures are smaller than
$T_e$(O~III).

{From} models computed with CLOUDY (Ferland 1996) it is found that $T_e$(Bac)
---labeled $T_e$(Hth) by CLOUDY--- is about 5\% smaller than
$T_e$(O~III). Moreover in the model with a homogeneous sphere of I~Zw~18 by
Stasinska and Schaerer (1999) it is found that $\langle T_e({\rm Ar~III})
\rangle = 16300$~K (for 63\% of the volume) and $\langle T_e ({\rm
Ar~IV})\rangle = 18300$~K (for 36\% of the volume) indicating the presence of
temperature variations. Notice that the average temperature has to be weighted
by the emissivities strengthening the effect of the temperature variations.
Moreover there is additional evidence that indicates that the temperature
variations are even higher than those predicted by photoionization models
(e. g. Peimbert 1995; Luridiana, Peimbert, \& Leitherer 1999; Stasinska and
Schaerer 1999). From equations (1--3) it follows that the smaller the adopted
temperature the smaller the derived He$^+$/H$^+$ value.

Due to the presence of very strong density variations inside gaseous nebulae
the different methods to derive the density yield very different values. The
root mean square density, $N_e$(rms), is usually obtained from the observed
flux in a Balmer line and by assuming a spherical geometry; $N_e$(rms) provides
a minimum value for $N_e$(He~II), the local density needed to derive the helium
abundance. Forbidden line ratios of lines of similar excitation energy give us
an average density for cases where the density is similar or smaller than the
critical density for collisional deexcitation; available line ratios in the
visual region are those of [S~II], [O~II], [Cl~III], and [Ar~IV], unfortunately
giant extragalactic H~II regions are close to the low density limit of these
ratios, the line intensities of [Cl~III] are very faint, for most observations
of the [O~II] lines the resolution is not high enough to separate $\lambda$3726
from $\lambda$3729 nor $\lambda$4711 of [Ar~IV] from $\lambda$4713 of
He~I. Consequently most of the densities in the literature are those derived
from the [S~II] lines, but as rightly mentioned by Izotov, Thuan, \& Lipovetsky
(1994, 1997) they are not representative of the regions where the He~I lines
originate. The self-consistent method, advocated by Izotov and collaborators,
is based only on line ratios of helium I and is the best method to derive the
density of the He~II zone: $N_e$(He~II).

The stellar underlying absorption can affect the derived He~I and H~I emission
line intensities and has to be taken into account. The best way to reduce this
effect is to avoid the presence of bright early type stars in the observed
slit, this can only be done for objects inside the Galaxy and the Magellanic
Clouds. To minimize the effect of the underlying absorption it is recommended
to use only objects where the line intensities show very large equivalent
widths in emission, and to increase the spectral resolution. The best way to
correct for underlying absorption is to use starburst models that predict the
underlying stellar spectrum. The correction for underlying absorption can be
tested by comparing the higher order Balmer lines, that are most affected by
this effect, with the brightest Balmer lines that are the least
affected. Similarly the underlying absorption effect is larger for
$\lambda$4471 and smaller for $\lambda\lambda$5876, 6678 and 7065; the effect
for these three lines is in general negligible due to a combination of causes
(mainly their large equivalent widths in emission).

\section{Ionization Structure}

The total He/H value is given by:

\begin{eqnarray}
\frac{N ({\rm He})}{N ({\rm H})} & = & \frac {N({\rm He}^0) + N({\rm He}^+) + 
N({\rm He}^{++})}{N({\rm H}^0) + N({\rm H}^+)},\\
& = & ICF({\rm He}) \frac {N({\rm He}^+) + N({\rm He}^{++})}
{N({\rm H}^+)}.                 
\end{eqnarray}
              
The He$^{++}$/H$^+$ ratio can be obtained directly from the 4686/H$\beta$
intensity ratio. In objects of low degree of ionization the presence of neutral
helium inside the H~II region is important and $ICF$(He) becomes larger than
1. The $ICF$(He) can be estimated by observing a given nebula at different
lines of sight since He$^0$ is expected to be located in the outer regions.
Another way to deal with this problem is to observe H~II regions of high degree
of ionization where the He$^0$ amount is expected to be negligible.
V\'{\i}lchez \& Pagel (1988) (see also Pagel et al. 1992) defined a radiation
softness parameter given by

\begin{equation}
\eta = \frac {N({\rm O}^+)N({\rm S}^{++})}{N({\rm S}^+)N({\rm O}^{++})};
\end{equation}                     

\noindent for large values of $\eta$ the amount of neutral helium is
significant, while for low values of $\eta$ it is negligible.

On the other hand for ionization bounded objects of very high degree of
ionization the amount of H$^0$ inside the He$^+$ Str\"omgren sphere becomes
significant and the $ICF$(He) can become smaller than 1. This possibility was
firstly mentioned by Shields (1974) and studied extensively by Armour et
al. (1999) for constant density models.  Since H$^0$ will be located in a thin
shell in the border of the nebula it can be shown that for ionization bounded
nebulae of homogeneous density the fraction of H$^0$/He$^+$ will be three times
larger for an observation of the whole object than for an observation of a line
of sight that includes the center.  Similarly this fraction becomes higher than
a factor of three for models with decreasing density from the center
outwards. Alternatively for density bounded nebulae this effect can be
neglected.

\section{SMC}

Peimbert, Peimbert, \& Ruiz (2000a) presented long slit observations of the
most luminous H~II region in the SMC: NGC~346. They divided the two long slit
positions into thirteen areas, four including the brightest stars ($m \sim 14$)
and 9 without stars brighter than $m = 17$. In the upper part of Figure 1 we
present a spectrum that includes all the observed areas (region 346--B); while
in the lower part of Figure 1 we present a spectrum made by the seven brightest
areas that do not include the brightest stars (region 346--A).

\begin{figure}
\plotfiddle{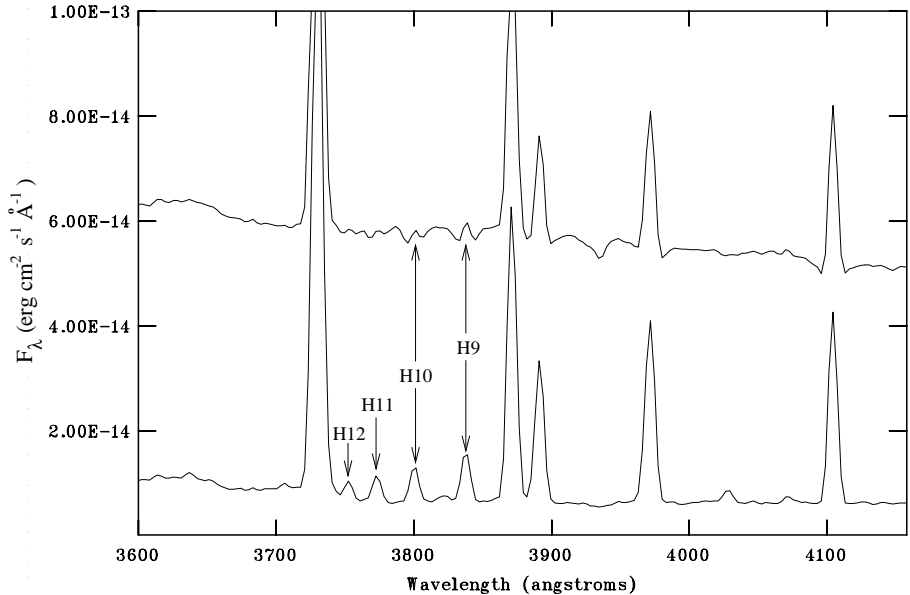}{220pt}{0}{80}{80}{-175}{-12}
\caption{Spectra of NGC~346 with and without underlying absorption. The 
vertical scale is
for the lower spectrum (region 346--A). The flux of the upper spectrum 
(region 346--B) was 
normalized to the
H$\alpha$ emission line flux of the lower spectrum.}
\end{figure}

After correcting region 346--A for extinction, based on the four brightest
Balmer lines, it is found that the weaker Balmer lines (H9 to H12) are not
affected by stellar underlying absorption (see Figure 1), therefore the He
lines are not expected to be affected by underlying absorption. The extinction
derived from the Balmer lines is in very good agreement with the extinction
derived from the stellar cluster by Massey, Parker, \& Garmany (1989), another
indication that the stellar underlying absorption of the Balmer lines is
negligible.

\begin{table}
\begin{center}
\caption{$N({\rm He}^+)/N({\rm H}^+)^a$ and $\chi^2$ for NGC~346}
\begin{tabular}{ccccccc}
\noalign{\smallskip}
\tableline
\noalign{\smallskip}
$T_e$(K) & \ & \multicolumn{5}{c}{$N_e$(cm$^{-3}$)}\\
\noalign{\smallskip}
&& 53 & 100 & 143 & 162 & 247\\
\noalign{\smallskip}
\tableline
\noalign{\smallskip}
11200 && 805    & 798    & 793    & 791    & {\bf 781}$^b$ \\
      && (83.2) & (47.7) & (26.4) & (20.0) & {\it (8.24)}$^c$ \\
\\
11800 && 806    & 799    & 793    & {\bf 790}    & 780 \\
      && (38.6) & (15.9) & (7.37) & {\it (6.59)} & (20.4) \\
\\
11950 && 806    & 799    & {\bf 793}        & 790    & 779 \\
      && (30.8) & (11.7) & {\bf (6.53)}$^d$ & (7.25) & (27.7) \\
\\
12400 && 807    & {\bf 799}    & 793    & 790    & 778 \\
      && (15.0) & {\it (7.17)} & (12.5) & (17.9) & (58.6) \\
\\
13000 && {\bf 809}    & 800    & 793    & 790    &  777 \\
      && {\it (9.72)} & (18.2) & (38.4) & (50.2) & (118) \\
\noalign{\smallskip}
\tableline
\tableline
\end{tabular}
\begin{list}{}{}
\item[$^a$]Given in units of $10^{-4}$, $\chi^2$ values in
parenthesis.
\item[$^b$]The He$^+$/H$^+$ values in boldface correspond to the minimum
$\chi^2$ values at a given temperature.
\item[$^c$]The minimum $\chi^2$ value at a given temperature is presented in
italics.
\item[$^d$]The smallest $\chi^2$ value for all temperatures and densities is
presented in boldface, thus defining $T_e$(He~II) and $N_e$(He~II).
\end{list}
\end{center}
\end{table}

To derive the He$^+$/H$^+$ value, in addition to the Balmer lines, we made use
of nine He~I lines, $\lambda\lambda$ 3889, 4026, 4387, 4471, 4921, 5876, 6678,
7065, and 7281 to determine $N_e$(He~II) and $T_e$(He~II) self-consistently. In
Table~1 we present He$^+$/H$^+$ values for different temperatures and
densities; the temperatures were selected to include $T_e$(O~III), $T_e$(Bac),
$T_e$(He~II), and two representative temperatures; the densities were selected
to include the minimum $\chi^2$ at each one of the five temperatures. The
temperature with the minimum $\chi^2$ is the self-consistent $T_e$(He~II) and
amounts to $11950\pm 560$~K; this temperature is in excellent agreement with
the temperature derived from the Balmer continuum that amounts to $11800 \pm
500$~K, alternatively $T_e$(O~III) amounts to $13070 \pm 100$~K. Notice that
the $\chi^2$ test requires a higher density for a lower temperature, increasing
the dependence on the temperature of the He$^+$/H$^+$ ratio. The values in
Table~1 correspond to the case where $\tau$~(3889) equals zero, for higher
values of $\tau$~(3889) the $\chi^2$ values increase. In Table~1 the
He$^+$/H$^+$ values in boldface and the $\chi^2$ values in italics correspond
to the minimum value of $\chi^2$ at a given $T_e$, the $\chi^2$ value in
boldface is the smallest value for all temperatures and all densities.

From Table~1 we obtain that $T_e$(He~II)$ = 11950$~K and $N_e$(He~II)$ =
143$~cm$^{-3}$, which correspond to He$^+$/H$^+ = 0.0793 \pm 0.007$. By
comparing the He/H values for lines of sight with different ionization degree
it is found that $ICF$(He) = 1.00. To obtain the total He/H value we have added
the contribution of He$^{++}$/H$^+$ that amounts to 2.2 x $10^{-4}$.

In Table~2 we present the helium abundance by mass $Y$(SMC), derived from
NGC~346. The $Y$(SMC) values were derived from Table~1, the He$^{++}$/H$^+$
values and the $Z$ value determined by Peimbert et al. (2000a).

\begin{table}
\begin{center}
\caption{$Y$(SMC)}
\begin{tabular}{ccccccc}
\noalign{\smallskip}
\tableline
\noalign{\smallskip}
$T_e$(K) & \ & \multicolumn{5}{c}{$N_e$(cm$^{-3}$)}\\
\noalign{\smallskip}
&& 53 & 100 & 143 & 162 & 247\\
\noalign{\smallskip}
\tableline
\noalign{\smallskip}
11200 && 0.2431 & 0.2416 & 0.2404 & 0.2399 & {\bf 0.2377}$^a$ \\
11800 && 0.2435 & 0.2419 & 0.2405 & {\bf 0.2399} & 0.2375 \\
11950 && 0.2436 & 0.2420 & {\bf 0.2405} & 0.2399 & 0.2374 \\
12400 && 0.2439 & {\bf 0.2421} & 0.2406 & 0.2399 & 0.2372 \\
13000 && {\bf 0.2443} & 0.2423 & 0.2407 & 0.2400 & 0.2370 \\ 
\noalign{\smallskip}
\tableline
\tableline
\end{tabular}
\begin{list}{}{}
\item[$^a$]Boldface values correspond to minimum $\chi^2$ values, see Table~1.
\end{list}
\end{center}
\end{table}

\section{$Y_p$}

To determine the $Y_p$ value from the SMC it is necessary to estimate
the fraction of helium present in the interstellar medium produced
by galactic chemical evolution. We will assume that
 
\begin{equation}
Y_p  =  Y({\rm SMC}) - Z({\rm SMC}) \frac{\Delta Y}{\Delta Z}.
\end{equation} 

Peimbert et al. (2000a) find that for $T_e$(He~II)$ = 11950 \pm 500$~K,
$Y$(SMC)$ = 0.2405 \pm 0.0018 $ and $Z$(SMC)$ = 0.00315 \pm 0.00063 $. To
estimate $\Delta Y/\Delta Z$ we will consider three observational
determinations and a few determinations predicted by chemical evolution models.

Peimbert, Torres-Peimbert, \& Ruiz (1992) and Esteban et al. (1999) found that
$Y = 0.2797 \pm 0.006$ and $Z = 0.0212 \pm 0.003$ for the Galactic H~II region
M17, where we have added 0.10dex and 0.08dex to the carbon and oxygen gaseous
abundances to take into account the fraction of these elements embedded in dust
grains (Esteban et al. 1998).  By comparing the $Y$ and $Z$ values of M17 with
those of NGC~346 (Peimbert et al. 2000a) we obtain $\Delta Y/\Delta Z = 2.17
\pm 0.4$. M17 is the best H~II region to determine the helium abundance because
among the brightest galactic H~II regions it is the one with the highest degree
of ionization and consequently with the smallest correction for the presence of
He$^0$ (i.e. $ICF$(He) is very close to unity). It can be argued that the M17
values are not representative of irregular galaxies, on the other hand they
provide the most accurate observational determination. {From} a group of 10
irregular and blue compact galaxies, that includes the LMC and the SMC, Carigi
et al.  (1995) found $\Delta Y/\Delta Z = 2.4 \pm 0.6$, where they added 0.2dex
to the O/H abundance ratios derived from the nebular data to take into account
the temperature structure of the H~II regions and the fraction of O embedded in
dust, moreover they also estimated that O constitutes 54\% of the $Z$
value. Izotov \& Thuan (1998) from a group of 45 supergiant H~II regions of low
metallicity derived a $\Delta Y/\Delta Z = 2.3 \pm 1.0$; we find from their
data that $\Delta Y/\Delta Z = 1.5 \pm 0.6$ by adding 0.2dex to the O
abundances to take into account the temperature structure of the H~II regions
and the fraction of O embedded in dust.

Based on their two-infall model for the chemical evolution of the Galaxy
Chiappini, Matteucci, \& Gratton (1997) find $\Delta Y/\Delta Z = 1.6$ for the
solar vicinity. Carigi (2000) computed chemical evolution models for the
Galactic disk, under an inside-out formation scenario, based on different
combinations of seven sets of stellar yields by different authors; the $\Delta
Y/\Delta Z$ spread predicted by her models is in the 1.2 to 1.9 range for the
Galactocentric distance of M17 (5.9 kpc).

Carigi et al. (1995), based on yields by Maeder (1992), computed closed box
models adequate for irregular galaxies, like the SMC, and obtained $\Delta
Y/\Delta Z = 1.52$. They also computed models with galactic outflows of well
mixed material, that yielded $\Delta Y/\Delta Z$ values similar to those of the
closed box models, and models with galactic outflows of O-rich material that
yielded values higher than 1.52. The maximum $\Delta Y/\Delta Z$ value that can
be obtained with models of O-rich outflows, without entering into contradiction
with the C/O and $(Z {\rm -C-O)/O}$ observational constraints, amounts to 1.69.

Carigi, Col\'{\i}n, \& Peimbert, (1999), based on yields by Woosley, Langer, \&
Weaver (1993) and Woosley \& Weaver (1995), computed chemical evolution models
for irregular galaxies also, like the SMC, and found very similar values for
closed box models with bursting star formation and constant star formation
rates that amounted to $\Delta Y/\Delta Z = 1.71$. The models with O-rich
outflows can increase the $\Delta Y/\Delta Z $, but they predict higher C/O
ratios than observed.

From the previous discussion it follows that $\Delta Y/\Delta Z = 1.9 \pm 0.5$
is a representative value for models and observations of irregular
galaxies. Moreover, this value is in good agreement with the models and
observed values of the disk of the Galaxy.

\begin{table}
\begin{center}
\caption{$Y_p$ Derived from the SMC}
\begin{tabular}{ccccccc}
\noalign{\smallskip}
\tableline
\noalign{\smallskip}
$T_e$(K) & \ & \multicolumn{5}{c}{$N_e$(cm$^{-3}$)}\\
\noalign{\smallskip}
&& 53 & 100 & 143 & 162 & 247\\
\noalign{\smallskip}
\tableline
\noalign{\smallskip}
11200 && 0.2363 & 0.2348 & 0.2336 & 0.2331 & {\bf 0.2309}$^a$ \\
11800 && 0.2373 & 0.2357 & 0.2343 & {\bf 0.2337} & 0.2313 \\
11950 && 0.2376 & 0.2360 & {\bf 0.2345} & 0.2339 & 0.2314 \\
12400 && 0.2384 & {\bf 0.2366} & 0.2351 & 0.2344 & 0.2317 \\
13000 && {\bf 0.2395} & 0.2375 & 0.2359 & 0.2352 & 0.2322 \\
\noalign{\smallskip}
\tableline
\tableline
\end{tabular}
\begin{list}{}{}
\item[$^a$]Boldface values correspond to minimum $\chi^2$ values, see Table~1.
\end{list}
\end{center}
\end{table}

The $Y_p$ values in Table~3 were computed by adopting $\Delta Y/\Delta Z = 1.9
\pm 0.5$ . The differences between Tables 2 and 3 depend on $T_e$ because the
lower the $T_e$ value the higher the $Z$ value for the SMC.

\section{Discussion}

The $Y_p$ value derived by us is significantly smaller than the value derived
by Izotov \& Thuan (1998) from the $Y$ -- O/H linear regression for a sample of
45 BCGs, and by Izotov et al. (1999) from the average for the two most metal
deficient galaxies known (I~Zw~18 and SBS 0335--052), that amount to $0.2443
\pm 0.0015$ and $0.2452 \pm 0.0015$ respectively.

The difference could be due to systematic effects in the abundance
determinations.  There are two systematic effects not considered by Izotov and
collaborators that we did take into account, the presence of H$^0$ inside the
He$^+$ region and the use of a lower temperature than that provided by the
[O~III] lines. We consider the first effect to be a minor one and the second to
be a mayor one but both should be estimated for each object.

From \ constant \ density \ chemicaly \ homogeneous \ models \ computed with
CLOUDY we estimate that the maximum temperature that should be used to
determine the helium abundance should be 5\% smaller than
$T_e$(O~III). Moreover, if there is additional energy injected to the H~II
region $T_e$(He~II) should be even smaller.

Luridiana, Peimbert, \& Leitherer (1999) produced a detailed photoionized model
of NGC~2363. For the slit used by Izotov, Thuan, \& Lipovetsky (1997) they find
an $ICF$(He) = 0.993; moreover they also find that the $T_e$(O~III) predicted
by the model is considerably smaller than observed.  {From} the data of Izotov
et al. (1997) for NGC~2363, adopting a $T_e$(He~II) 10\% smaller than
$T_e$(O~III) and $\Delta Y/\Delta Z = 1.9 \pm 0.5$ we find that $Y_p = 0.234
\pm 0.006$.

Similarly, Stasinska \& Schaerer (1999) produced a detailed model of I~Zw~18
and find that the photoionized model predicts a $T_e$(O~III) value 15\% smaller
than observed, on the other hand their model predicts an $ICF$(He) = 1.00.
From the observations of $\lambda\lambda$ 5876 and 6678 by Izotov et al.
(1999) of I~Zw~18, and adopting a $T_e$(He~II) 10\% smaller than $T_e$(O~III)
we obtain $Y_p = 0.237 \pm 0.007$; for a $T_e$(He~II) 15\% smaller than
$T_e$(O~III) we obtain $Y_p = 0.234 \pm 0.007$, both results in good agreement
with our determination based on the SMC. Further discussion of these issues is
presented elsewhere (Peimbert, Peimbert, \& Luridiana 2000b).

The primordial helium abundance by mass of $0.2345 \pm 0.0030 (1 \sigma)$ ---
based on the SMC --- combined with the computations by Copi, Schramm, \& Turner
(1995) for three light neutrino species implies that, at the 95 percent
confidence level, $\Omega_b h^2$ is in the 0.0046 to 0.0103 range.  For $h =
0.65$ the $Y_p$ value corresponds to $0.011 < \Omega_b < 0.024$, a value
considerably smaller than that derived from the pregalactic deuterium
abundance, $D_p$, determined by Burles \& Tytler (1998) that corresponds to
$0.040 < \Omega_b < 0.050$ for $h$ = 0.65, but in very good agreement with the
observational estimate of the global budget of baryons by Fukugita, Hogan, \&
Peebles (1998) who find $0.007 < \Omega_b < 0.038$ for $h$ = 0.65. The
discrepancy between $Y_p$ and $D_p$ needs to be studied further.

To increase the accuracy of the $Y_p$ determinations we need observations of
very high quality of as many He~I lines as possible to derive $T_e$(He~II),
$N_e$(He~II), and $\tau$(3889) self-consistently. We also need observations
with high spatial resolution to estimate the $ICF$(He) along different lines of
sight.

\bigskip

It is a pleasure to acknowledge several fruitful discussions on this subject
with: L. Carigi, V. Luridiana, B. E. J. Pagel, M. T. Ruiz, E. Skillman,
G. Steigman, S. Torres-Peimbert, and S. Viegas.

\end{document}